\documentclass[preprint,authoryear,11pt,3p]{elsarticle}

\usepackage{setspace,mathtools,amssymb,siunitx,booktabs,kpfonts,xcolor}
\usepackage[font=itshape]{quoting}
\onehalfspace

\usepackage{amssymb,multirow,amsmath,graphicx,graphics,array,lscape,rotating,color,amsthm,ifsym, colortbl, courier,listings,enumitem,hyperref,wrapfig}
\usepackage{algorithm}
\usepackage{comment}
\usepackage{algpseudocode}
\usepackage{bm}
\definecolor{dkgreen}{rgb}{0,0.6,0}
\definecolor{lightgreen}{rgb}{0.7,0.95,0.7}
\definecolor{gray}{rgb}{0.9,0.9,0.9}
\definecolor{mauve}{rgb}{0.58,0,0.82}
\definecolor{lightred}{rgb}{0.95,0.8,0.8}
\definecolor{lblue}{rgb}{0.2,0.45,0.6}
\lstdefinestyle{mystyle}{frame=tb,
  language=R,
  aboveskip=3mm,
  belowskip=3mm,
  showstringspaces=false,
  columns=flexible,
  basicstyle={\footnotesize\ttfamily},
  numbers=none,
  numberstyle=\tiny\color{gray},
  keywordstyle=\color{blue},
  commentstyle=\color{dkgreen},
  stringstyle=\color{mauve},
  breaklines=true,
  breakatwhitespace=true,
  tabsize=3
}
\usepackage{bbm}
\usepackage{makecell}

\newcommand{\cmark}{\ding{51}}%
\newcommand{\xmark}{\ding{55}}%

\begin{document}

\begin{frontmatter}

\title{The future of forecasting competitions: \\ Design attributes and principles}

\author[add1]{Spyros Makridakis}
\ead{makridakis.s@unic.ac.cy}
\author[add2]{Chris Fry}
\ead{chrisfry@google.com}
\author[add3]{Fotios Petropoulos}
\ead{f.petropoulos@bath.ac.uk}
\author[add4]{Evangelos Spiliotis}
\ead{spiliotis@fsu.gr}
\cortext[]{Correspondence: Fotios Petropoulos, School of Management, School of Management, University of Bath, Claverton Down, Bath, BA2 7AY, UK.}

\address[add1]{Institute for the Future (IFF), University of Nicosia, Nicosia, Cyprus}
\address[add2]{Google Inc., USA}
\address[add3]{School of Management, University of Bath, UK}
\address[add4]{Forecasting and Strategy Unit, School of Electrical and Computer Engineering, \\National Technical University of Athens, Greece}

\begin{abstract}

Forecasting competitions are the equivalent of laboratory experimentation widely used in physical and life sciences. They provide useful, objective information to improve the theory and practice of forecasting, advancing the field, expanding its usage and enhancing its value to decision and policymakers. We describe ten design attributes to be considered when organizing forecasting competitions, taking into account trade-offs between optimal choices and practical concerns like costs, as well as the time and effort required to participate in them. Consequently, we map all major past competitions in respect to their design attributes, identifying similarities and differences between them, as well as design gaps, and making suggestions about the principles to be included in future competitions, putting a particular emphasis on learning as much as possible from their implementation in order to help improve forecasting accuracy and uncertainty. We discuss that the task of forecasting often presents a multitude of challenges that can be difficult to be captured in a single forecasting contest. To assess the caliber of a forecaster, we, therefore, propose that organizers of future competitions consider a multi-contest approach. We suggest the idea of a forecasting ``athlon'', where different challenges of varying characteristics take place.

\end{abstract}

\begin{keyword}
data science \sep business analytics \sep competitions \sep organization \sep design \sep forecasting
\end{keyword}

\end{frontmatter}

\clearpage

\section{Introduction}

\cite{Hyndman2020-gs} reviewed the history of time series forecasting competitions and discusses what we have learned from them as well as how they have influenced the theory and practice of forecasting. In this article, we provide a systematic approach to the design of forecasting competitions, focusing on forecasting competitions that allow participants to submit their forecasts, thus excluding early studies where all the methods and approaches were provided by the researchers \citep[see, for example,][]{Newbold1974-yk,Makridakis1979-lv}. 

In this sense, the first time series forecasting competition, M \citep{Makridakis1982-co}, was held in 1981 with seven participants, all known to the organizer and invited personally by telephone or regular mail, while the M5 \citep{Makridakis2020-wq, Makridakis2020-bj}, hosted by Kaggle in 2020 and run over the internet, attracted in its two tracks (accuracy and uncertainty) 8,229 participants from 101 countries around the world, offering \$100,000 prizes to the winners. There is little doubt, therefore, that forecasting competitions have changed a great deal and have become big events, attracting large numbers of participants from diverse backgrounds and with varying reasons to join. As time passes, however, there is a need to question the way forecasting competitions are structured, and consider improvements in their design and the objectives they strive to achieve in order to attain maximum benefits from their implementation. Also, as presented in the encyclopedic overview of \cite{Petropoulos2020-ec}, the applications of forecasting expand to many social science areas, such as economics, finance, health care, climate, sports, and politics, among others. As such, there is also the need to consider new application areas for future forecasting competitions beyond operations, supply chain, and energy, which have been the main case till now.

Forecasting competitions are the equivalent of the laboratory experimentation widely used in physical and life sciences. They are used to evaluate the forecasting performance of various approaches and determine their accuracy and uncertainty. Their broad purpose is to provide objective, empirical evidence to aid policy and decision makers about the most appropriate forecasting approach to use to realize their specific needs.

There have been many commentaries over time on the design and limitations of such competitions (see, for instance, discussions and commentaries of issues 17:4 and 36:1 of the \textit{International Journal of Forecasting} for the case of the M3 and M4 forecasting competitions). However, given the large number of forecasting competitions conducted over the last decade, organized from both academic teams but also companies and organizations, a structured analysis of their design attributes seems to be necessary. Moreover, we deliberate about the future of forecasting competitions and what should be done to improve their value and expand their usefulness across application domains. In this regard, we provide a systematic review of past forecasting competitions, determining their main attributes and key innovations, while also proposing how future competitions could be designed so that we better learn from data and ``data analysis'' \citep{Donoho1384734} of the competitions' results, the circumstances under which a forecasting method is expected to work best, instead of just focusing on the winners. 

The paper consists of six sections and a conclusion. After this short introduction, section \ref{sec2} summarizes the conclusions of Hyndman's influential paper about past time series forecasting competitions, an interest of the present discussion, and enumerates his suggestions about the characteristics of future ones. Section \ref{sec3} describes various types of forecasting competitions, considering their scope, the type of data used in terms of diversity and representativeness, structure, granularity, availability, the length of forecasting horizon and several other attributes, including performance measures and the need for benchmarks. Consequently, section \ref{sec4} identifies the commonalities as well as design gaps of past forecasting competitions by mapping the designs of indicative, major ones to the attributes described previously and mentioning the advantages and drawbacks of each. Section \ref{sec5} focuses on outlining the proposed features of some ``ideal'' forecasting competitions that would avoid the problems of past ones while filling existing gaps in order to improve their value and gain maximum benefits from their implementation. Section \ref{sec6} presents some thoughts about institutionalizing the practice of forecasting competitions and systematizing the way they are conducted, moving from running single competitions to structuring them across multiple forecasting challenges in the way that pentathlons are run with single winners in each challenge and an overall one across all. Finally, the conclusion summarizes the paper and proposes expanding the competitions beyond business forecasting to cover other social science areas in need of objective information to improve policy and decision making. 

\section{A brief history of time series forecasting competitions}
\label{sec2}

In his paper, \cite{Hyndman2020-gs} concludes that time series forecasting competitions have played an important role in advancing our knowledge of what forecasting methods work and how their performance is affected by various influencing factors. He believes that in order to improve objectivity and replicability, the data and the submitted forecasts of competitions must be made publicly available in order to promote research and facilitate the diffusion and the usage of their findings in practice. At the same time, their objectives must be clear and the extent that their findings can be generalized must be stated. According to him, future competitions should carefully define the population of data from which the sample has been drawn and the possible limitations of generalizing their findings to other situations. The usage of instance spaces \citep{Kang2017-ha} could provide a way to specify the characteristics of the data included and allow comparisons to other competitions or data sets with well known properties \citep{Fry2020-ye,Spiliotis2020-gl}. Moreover, a nice side-effect of time series competitions is that they have introduced popular benchmarks, allowing the evaluation of performance improvements and comparisons among competitions for judging the accuracy and uncertainty of the submitted methods, including the assessment and replication of their findings over time. Furthermore, as new competitions emerge and the benchmarks are  regularly  updated, the effect of developing methods that overfit published data is mitigated and new, robust forecasting methods can be effectively identified.

On the negative side, Hyndman expresses concerns about the performance measures used, stating that these should be based on well-recognized attributes of the forecast distribution. This is particularly true for the case of the prediction intervals, stating that the widely used Winkler scores \citep{Winkler1972-fe} are not scale-free and that their scaled version used to assess the interval performance in the M4 competition \citep{Makridakis2020-mm} seems rather ad hoc, with unknown properties. Consequently, he cites the work of \cite{Askanazi2018-no} who assert that comparisons of interval predictions are problematic in several ways and should be abandoned for density forecasts. Probabilistic forecasts, such as densities, could be evaluated instead using proper scoring rules and scale-free measures like log density scores, as done in M5 and in some energy competitions \citep{Hong2016-my,Hong2019-ay}. There is, therefore, a need to reconsider how such probabilistic forecasts will be made and evaluated in future competitions to avoid the criticisms that they are inadequate. However, no matter how such evaluations are done, Hyndman suggests that it would be desirable that forecast distributions will be part of all future forecasting competitions. Another issue he raises is whether explanatory/exogenous variables improve forecasting performance over that of time series methods. For instance, in the tourism forecasting competition \citep{Athanasopoulos2011-lw} explanatory/exogenous variables were helpful only for one-step-ahead forecasts, while in some energy competitions \citep{Hong2014-sz,Hong2016-my,Hong2019-ay} using temperature forecasts was beneficial for short-term forecasting, where weather forecasts were relatively accurate, with the results being mixed for longer forecasting horizons. On the other hand, explanatory/exogenous variables whose values can be specified, such as existence of promotions, day of the week, holidays, and days of special events like the super bowl, are generally considered being helpful for improving forecasting performance and should be therefore included in the forecasting process \citep{Makridakis2020-eu}.

A major suggestion of \cite{Hyndman2020-gs}, previously discussed by the commentators of the M3 competition \citep{Fildes2001comment,Hyndman2001comment}, is that future time series competitions should focus more on the conditions under which different methods work well rather than simply identifying the methods that perform better than others. Doing so will present a significant change that will be particularly relevant for breaking the black box of machine and deep learning forecasting methods that will be necessary to better understand how their predictions are made and how they can be improved by concentrating on the factors that influence accuracy and uncertainty the most. In addition, he believes that future time series competitions should involve large-scale multivariate forecasting challenges while focusing on irregularly spaced and high frequency series such as hourly, daily, and weekly data that is nowadays widely recorded by sensors, systems, and the internet of things. Finally, Hyndman states that he does not know of any large-scale time series forecasting competition that has been conducted using finance data (e.g., stock and commodity prices and/or returns) and that such a competition would seem to be of great potential interest to the financial industries and investors in general.

\section{Design attributes of forecasting competitions}
\label{sec3}

In this section we identify and discuss ten key attributes that should be considered when designing forecasting competitions, even if some of them might not be applicable to all of them. Table 1 provides a summary description of these attributes, which are then discussed in detail in the next subsections.

\begin{table*}[ht]
\small
\centering
  \caption{Summary description of the design attributes of forecasting competitions.}
    \begin{tabular}{p{0.022\linewidth}  p{0.33\linewidth}  p{0.55\linewidth}}
    \hline
 & \textbf{Design attribute} & \textbf{Description} \\
\hline
1 & Scope & Focus of competition (domain or application); Type of submission  (numeric or judgment); Format of submission (point forecasts, uncertainty estimates, or decisions). \\
\hline
2 & Diversity and representativeness & Degree to which the findings and insights obtained can be generalized and applied to other settings or data sets. \\
\hline
3 & Data structure & Degree the data are connected or related; Explanatory or exogenous variables used for supporting the forecasting process. \\
\hline
4 & Data granularity & The most disaggregated level, cross-sectional or temporal,  where data are available. \\
\hline
5 & Data availability & Amount of information provided by the organizers for producing the requested forecasts (e.g. number of series contained in the data set and historical observations available per series). \\
\hline
6 & Forecasting horizon & Length of time into the future for which forecasts are requested. \\
\hline
7 & Evaluation setup & Number of evaluation rounds  (single vs. rolling origin); Live vs. concealing data competitions. \\
\hline
8 & Performance measurement & Measures used for evaluating performance in terms of forecasting accuracy/uncertainty, utility, or cost. \\
\hline
9 & Benchmarks & Standards of comparisons used for assessing performance improvements. \\
\hline
10 & Learning & What can be learned for advancing the theory and practice of forecasting; Replicability or reproducibility of results; Making and evaluating hypotheses about the findings of the competitions; Challenging and confirming the results of past competitions.\\
\hline
    \end{tabular}
  \label{tab:attributes}
\end{table*}

\subsection{Scope}\label{sec3-1}
The first decision in designing a forecasting competition relates to its scope, which can be defined based on (\textit{i}) the focus of the competition, (\textit{ii}) the type of the submissions it will attract, and (\textit{iii}) the format of the required submissions.

Regarding the focus, there is a spectrum of possibilities ranging from generic to specific competitions. Generic competitions feature data from multiple domains that represent various industries and applications, as well as from various frequencies. Examples include the M, M3, and M4 forecasting competitions that include data from different domains (micro, macro, industry, demographic, finance, and others) and various frequencies (yearly, quarterly, monthly, weekly, daily, hourly and others). While the results of such competitions identify the methods performing best on each data domain/frequency, they typically determine the winners based on their average performance across the complete data set. Thus, although their main findings may not be necessarily applicable to all the domains/frequencies examined, they help us effectively identify best forecasting practices that hold for diverse types of data.

Specific competitions feature data of a particular domain/frequency, a particular industry or company/organization. Examples of such competitions include the global energy ones and the majority of those hosted on Kaggle \citep{Bojer2020-mc}, including M5. Although these competitions may be more valuable for specific industries or organizations, replicating real-world situations, their findings are restricted to the specific data set and cannot be generalized to other situations. Finally, semi-specific competitions feature data that although refers to a particular domain, it includes instances from various applications of that domain which may therefore require the utilization of significantly different forecasting methods. For example, a semi-specific energy competition may require forecasts for renewable energy production, energy demand, and energy prices, with the winners being determined based on their average performance across these tasks. In this case, factors that influence forecasting performance in the examined domain can be effectively identified while the key findings of the competitions can  be applicable to several  forecasting tasks of that domain.

Apart from the focus, when deciding on the scope of a competition, organizers will need to think about the types of submissions that they would receive, particularly if these submissions will be based on automatic statistical algorithms or human judgment. While most competitions do not state this explicitly, the type of submissions is usually implied based on the number of inputs required. In a large-scale forecasting competition, where one has to provide many thousands of inputs, automatic algorithms might be the only feasible way. In smaller scale competitions, judgment could be used in predicting events while in cases where data is insufficient or even unavailable, judgment may be the only possible way to produce forecasts and estimate uncertainty. Consider, for instance, challenges similar to the ones posed within the Good Judgment project\footnote{https://goodjudgment.com/} and questions such as “what is the possibility that humans will visit Mars before the end of 2030?”. In such cases, the focus of the competition will be the events examined and the required submissions will have to be made judgmentally.

A third decision on the scope of a competition has to do with the format of the submissions requested from the participants. While some of the forecasting competitions so far have asked for the submission of point forecasts only, it is preferable that submissions of uncertainty should be required too. This can be obtained by the submission of prediction intervals for one or multiple indicative quantiles or, even better, the submission of a fine grid of quantiles including the extreme tails. If the event to be forecast has a discrete number of possible solutions, uncertainty can be provided in a form of confidence levels (e.g., 90\% certainty) or categorical answers (e.g., low, moderate, and high confidence). Note also that it can be the case that a forecasting competition does not ask for forecasts (or estimates of uncertainty) per se, but the decisions to be made when using such forecasts. Examples include setting the safety stock in an inventory system, the selection of a portfolio of stocks in investing, or betting amounts for future events given their odds. 

Finally, we believe that there is a need to have clear objectives and hypotheses before the commencing of a forecasting competition and define its scope-related attributes accordingly. Recent forecasting competitions have followed this example \citep[see, for instance,][]{MAKRIDAKIS202029}, thus avoiding the problem of HARKing \citep{Kerr1998-uv,Lishner2021-wc} by rationalizing the findings after the fact in perfect foresight (hindsight bias) or being driven by findings which are directly biased by the design of the competition itself. This is standard practice in other fields and is closely linked with the promotion of open research and the avoidance of ``$p$-hacking''. For example, psychological studies are often pre-registered, an increasingly popular requirement for many academic journals.

\subsection{Diversity and representativeness}\label{sec3-2}
Regardless if the focus of a competition is generic or not, it is important that the events considered have a reasonable degree of diversity that will allow for generalization of the findings and insights obtained. Diversity effectively refers to the heterogeneity of the events to be predicted. In the case of the forecasting competitions that provide historical information in the  form of time series, diversity is usually determined by visualizing spaces based on time series features \citep{Kang2017-ha} that may include the strength of predictable patterns (trend, seasonality, autocorrelations, etc.), the degree of predictability (coefficient of variation, signal-to-noise ratio, entropy, etc.), the degree of intermittence and sparseness (fast versus slow-moving items), as well as the length and periodicity of the data, among others. In time series such features can be endogenously measured while in competitions where past data is not provided, diversity can be appreciated with regards to the intent of the events under investigation and the implicit requirements from a participant’s perspective in analyzing and producing forecasts/uncertainty for such events. 

Diversity could also include the country of origin of the data, the type of data domains, the frequencies considered, the industries or companies investigated, and the time frame covered. For example, the results of a competition like M5 that focused on the sales of ten US stores from a global grocery in 2016 would not necessarily apply to a grocery retailer in China in the same year or another United States grocery retailer in 2021. Similarly, they may not apply to other types of retailers, such as fashion, pharmaceutical, or technology, to firms operating online, or providing different discounts and promotions strategies. Diversifying the data set of the competition so that multiple events of different attributes are considered is a prerequisite for designing competitions to represent reality realistically and ensure that its findings can be safely generalized across the domain(s), frequencies, or application(s) being considered.

Other competitions could be based on forecasting data of unknown or undisclosed sources, as well as competitions based on forecasting synthetic time series (i.e., time series data generated through simulations). Such competitions would allow identifying the conditions where particular forecasting models perform well, including time series characteristics, such as seasonality, trend, noise, and structural changes, as well as decisions like the forecasting horizon considered. These competitions would enable learning more from their results, understanding how methods and models that obey particular theoretical properties and assume certain distributions would work under real-life empirical settings.

\subsection{Data structure}\label{sec3-3}
While in some competition settings it is possible that no data is provided at all, in most competitions historical data is made available. Such data may be individual time series that are not somehow connected to one another. In such cases, although series are typically forecast separately, participants may attempt to apply cross-learning techniques to improve the accuracy of their solutions, as was the case with the two top-performing solutions in the M4 competition \citep[see for example][]{MONTEROMANSO202086, Semenoglou2020-mq,monteromanso2021principles}. It is also possible that competition data is logically organized to form hierarchical structures \citep{Hyndman2011-jm}. Such structures do not have to be uniquely defined necessarily. For instance, in competitions like M5, the sales of a company may be disaggregated by regions, categories, or both if grouped hierarchies are assumed. Given that in many forecasting applications hierarchies are present and information exchange between the series is possible, deciding on the correlation of the data provided is critical for determining under which circumstances the findings of the competition will apply. 

Alternatively, the provided time series data may or may not be supported by additional information. For example, in competitions like M4 where the existence of timestamps may have led to information leakage about the actual future values of the series, dates should not be provided. However, when this information is indeed available, then multivariate settings may also be considered. Also, while data availability might be limited to the variables for which forecasts are required, explanatory/exogenous variables can also be provided. Information for such variables may match the time window of the dependent variables, part of it, or even exceed it. Explanatory/exogenous variables may be either provided by the organizers of the competition to its participants directly or collected by them through various external sources. In any case, it is important that the explanatory/exogenous variables used for producing the forecasts will only refer to information that would have been originally available at the time the forecasts were produced and not after that point to make sure that no information about the actual future is leaked. For example, short-term weather forecasts may be offered as an explanatory variable for predicting wind production, but no actual future weather conditions measured either on site or at a nearby meteorological station.

\subsection{Data granularity}\label{sec3-4}
Data granularity refers to the most disaggregated level where data will be available and may refer both to cross-sectional and temporal aggregation levels \citep{Spiliotis2020-hj}. In most cases, the granularity of the data matches that of the variable to be forecast, but this does not have to always be the case. If, for example, a competition focuses on the sales of a particular product in the European Union, then country-level sales or even store-level sales might be helpful in improving forecasting performance. Similarly, smart-meter data may enhance the predictions of energy consumption at city level, with hourly measurements being also useful in predicting daily demand. This is particularly true in applications where data appear in mixed frequencies. For instance, in econometric regression, a quarterly time series may be used as an external regressor in forecasting a monthly time series. 

Temporal granularity is more relevant when the data under investigation is organized over time (time series data). Increasingly, forecasting competitions have been focusing on higher frequency data like daily and weekly series, but this should not be considered a panacea for all future competitions. The choice of the frequency needs to be linked with the scope of the competition as low-frequency data will be naturally used for supporting strategic decisions, while low-frequency ones for supporting operations. For instance, daily data are not available for macroeconomic variables compared to monthly, quarterly, or yearly frequencies. Similarly, daily or hourly data would be more relevant in forecasting the sales of fresh products for store replenishment purposes. Finally, special treatment should be given in instances where seasonality is not an integer number as, for example, when using weekly frequency data. 

\subsection{Data availability}\label{sec3-5}
Data availability refers to the amount of information provided by the organizers for producing the requested forecasts. For time series competitions this would include the number of historical observations available per series as well as the number of series contained in the data set. Note that both dimensions of data availability may be equally important in determining the performance of the submitted forecasts. For instance, in time series competitions, methods can be trained both in a series-by-series fashion, where a large number of historical observations is desirable per series, and in a cross-learning one, where data sets of multiple series are preferable for building appropriate models. In general, relatively large data sets are more advantageous over smaller ones so that the participants will be capable of effectively training their models by extracting more information from the data. In addition, the probability of a participant winning the competition by luck rather than skills is effectively reduced. For example, in competitions of the size of the M4, which involved 100,000 series, it is practically impossible to win by making random choices \citep{Spiliotis2020-gl}.

Data availability can be also driven by the scope of the competition and the type of the events to be predicted. For example, if the competition focuses on new product or technological forecasting, data availability will be naturally limited over time. Similarly, if the competition focuses on the sales of a manufacturer that produces a limited number of products, data availability will be naturally bounded over series, requiring more manufactures of the same industry to be included in the data set to expand its size and improve its representativeness. Moreover, data availability may be influenced by the frequency of the series, especially when multiple periodicities are observed. Hourly electricity consumption data, for instance, may contain three seasonal cycles: daily (every 24 hours), weekly (every 168 hours), and yearly (every 12 months). In addition, when dealing with seasonal data, it is generally believed that a minimum of three seasonal periods are required in order for the seasonal component of the series to capture the periodic patterns existing across time. 

Certain domain-specific future forecasting competitions may not offer any data at all. In the era of big data and instant access to many publicly available sources of information, participants are usually in a position to gather the required data by themselves, but also to complement their forecasts by using any other publicly available information. However, in the case that the organizers decide not to provide data, there is still a benefit to specifying a ``default'' data set to be used for evaluation purposes. Finally, in non-time-series forecasting competitions, such as the Good Judgment project, quantitative data may not only not be provided but it may not be available at all.

\subsection{Forecasting horizon}\label{sec3-6}
The forecasting horizon may vary from predicting the present situation (also known as nowcasting, especially popular in predicting macroeconomic variables), to immediate, short, medium, and long-term planning horizons. The exact definition of each planning horizon may differ with regards to the frequency of the data under investigation. For instance, for hourly data, 1-24 hours ahead is usually considered short-term forecasting. At the same time, 1-3 months ahead can also be regarded as ``short-term'' when working with monthly data. Accordingly, the forecasting horizon can be naturally bounded based on the frequency of the series. For daily data, for example, it is probably unreasonable to produce forecasts for the following three years, a request which is reasonable for quarterly data.

The choice of the appropriate forecasting horizon is a function of various factors that may include the importance of the specific planning horizons for the application data, the user of the forecasts, and the hierarchical level of the forecast. Short-term horizons are suitable for operational planning and scheduling, mid-term horizons are appropriate for financial, budgeting, marketing, and employment decisions, while long-term forecasts are associated with strategic decisions that include technological predictions as well as business and capacity planning. 

It is not uncommon in forecasting competitions to require forecasts for multiple periods ahead, with the performance usually being measured as the average across all horizons. However, for some applications, like store replenishment or production, it is more relevant to consider the cumulative forecast error (difference between the sum of actual values and the sum of the forecasts for the lead time) rather than the average of the forecast errors across all horizons. In other applications, specific forecast horizonsmay be more important than others, so averaging across all horizons may not be so useful.

\subsection{Evaluation setup}\label{sec3-7}
In time-series forecasting competitions, the most common design setup is to use historical data and conceal part of it to be used as test data to evaluate the performance of the submitted forecasts. The setup of concealing data may be further expanded to a number of rolling evaluation rounds. In single origin evaluation, participants do not receive feedback on their performance which is based on a single time window, which may not be representative of the entire series. For example, in electricity load forecasting where three strong seasonal patterns are typically observed across the year, evaluating submissions by considering only one particular day, week, or month is not appropriate. Similarly, we found this to be a drawback of the evaluation setup used in M5.

To avoid the disadvantage of a single origin, the competition can be rolling \citep{Tashman2000-mv}, revealing some more of the hidden data each time and asking for new forecasts at each rolling iteration, providing the participants the opportunity to learn and improve their performance over time. A potential disadvantage of rolling-origin competitions is that they require more inputs and energy by the participants who may wish, or have to adjust their models at each new round. For this reason, rolling origin competitions display higher drop-out rates, excluding also participants that are interested in participating but missed some early rounds and those that cannot be committed for a long period of time. An alternative could be a rolling-origin evaluation set-up where the participants provide the code for their solutions, and then the organizers produce forecasts automatically for multiple origins, as required. Yet, even if a forecasting competition does not have a rolling-origin evaluation design, participants may still decide to perform rolling-origin evaluation on the available (not the concealed) data to develop their algorithms, validate their performance under different settings, and select proper hyper-parameters. This is closely related to the concept of time series cross-validation.

Instead of concealing data, a competition can be designed to take place on a real-time basis (live competition), with forecasts being evaluated against the actual data once they become available. The major advantage of live competitions is that participants can incorporate current information to their forecasts in real time, meaning that data and external variables could be fetched by the participants themselves based on their preferences and methods used. Also, information leakage about the actual future values becomes impossible and the competition represents reality perfectly. Its disadvantage is that it is much more difficult to run (e.g., data must be collected in real time and evaluations must be accordingly updated) while taking some time until the actual values become available. A real-time competition may have a single submission origin or multiple, rolling ones. In the latter case, feedback is explicitly provided to the participants in real time, allowing learning with each additional rolling iteration. Its major disadvantage is that it would be much more difficult to run and would require great motivation to participate given the considerable effort to keep informed and update the forecasts each time.  

In some cases, when historical information is not available, concealing data is not an option. In such cases, the real-time design is the only alternative. Examples include elections and sports forecasting, where a single evaluation origin will typically be possible. However, participants may be also allowed to submit multiple forecasts (or revise previously submitted forecasts) until a particular point in time in live submission setups that include, for instance, prediction markets.

\subsection{Performance measurement}\label{sec3-8}
Another important decision in designing a competition is how the performance will be measured and evaluated. It is common that the performance of the (point) forecasts is evaluated using statistical error measures. The choice of such measures should be based on a variety of factors, such as their theoretical foundation, applicability, and interpretability. Nowadays, relative and scaled error measures are generally preferred to percentage ones \citep{Hyndman2006-rp}, however the latter are still dominant in practice by being more intuitive. The evaluation of the estimation of the uncertainty around the forecasts can be performed using interval scores and proper scoring rules \citep{Makridakis2020-bj}. Proper scoring rules can address both sharpness and calibration, which is relevant in estimating the performance under fat tails. In all cases, however, robust measures, with well-known statistical properties should be preferred to interpret the results and be confident of their value.

In cases where the importance (volume and value) of the predicted events varies, performance measurements may include weighting schemes that account for such differences. This is especially true when evaluating the performance of hierarchical structured data where some aggregation levels may be more important than others based on the decisions that the forecasts will support. For instance, product-store forecasts may be considered more important than regional ones when used for supply-chain management purposes, with the opposite being true in cases where forecasts are used for budgeting purposes. Similarly, forecasts that refer to more expensive or perishable products may be weighted more than those that refer to inexpensive, fast-moving ones.

Whenever possible, instead of measuring the performance of the forecasts, one should measure their utility value directly. For instance, if the forecasts referred to investment decisions, the actual profit/loss from such investments could be measured. If the forecasts were to be used in a supply-chain setting, then inventory-related costs, achieved service levels, and/or the variance of the forecasted variable can be useful measurements of their utility \citep{Petropoulos2019-bu}. If more than two performance indicators need to be considered, then multicriteria techniques could be used to balance the performance across the chosen criteria. A simpler approach would be to assume equal importance across criteria and apply a root mean square evaluation measure. Care should be used to address any double-counting that can arise when evaluating hierarchical series with multiple related levels.

Another critical factor in evaluating forecasts is the cost relating to various functions of the forecasting process, including data collection, computational resources required to produce the forecasts \citep{Nikolopoulos2018-wa}, and personnel time that is needed to revise/finalize such forecasts when judgment is needed. In standard forecasting competitions where data is provided and the submission format usually refers to automatic forecasts, the computational cost can be easily measured by sharing the code used for their production and reproducing them. Once the computational cost is determined, it is important to contrast any improvements in performance against any additional costs. Effectively, this becomes a Forecast Value Added (FVA) exercise \citep{Gilliland2013-dw,Gilliland2019-xl}, accepting that computational time is often subject to programming skills and optimizations techniques, making its correct estimate a considerable challenge.

\subsection{Benchmarks}\label{sec3-9}
A important decision in designing competitions similar to selecting the performance measurements has to do with the choice of appropriate benchmarks. Such benchmarks should include both traditional and state-of-the-art models and algorithms that are suitable for the competition based on its scope and particularities of the data. Usually, benchmarks include individual methods that have performed well in previous, similar competitions, are considered standard approaches for the forecasting task at hand, or display a performance which is considered a minimum for such a task. For example, ARIMAX, linear regression, and decision-tree-based models can be used as benchmarks in competitions that involve explanatory/exogenous variables, Croston's method in competitions that refer to inventory forecasting, the winning methods of the first three M competitions for the fourth one, and a random walk model for the performance of a major index such as S\&P 500 or FTSE in a stock market competition. Simple combinations of state-of-the-art methods are also useful benchmarks, especially given the ample evidence on their competitive performance \citep{Makridakis2020-eu}. It is good practice that the implementation of the benchmark methods is fully specified. This will allow participants to obtain a valid starting point for their investigation and facilitate transparency and reproducibility, indicating the additional value-added of a proposed method over that of an appropriate benchmark.

\subsection{Learning}\label{sec3-10}
Regardless of the design of the competition, its objective should not be just to determine the winners of the examined forecasting task, but also learn how to advance the theory and practice of forecasting by identifying the factors that contribute to the improvement of the forecasting accuracy and the estimation of uncertainty. This has been the case for the competitions organized by academics, but not in all others. 
In order to allow for such learning, sufficient information would be required about how the forecasts are made by the participants, with the code used (where applicable) being also published to facilitate replicability or reproducibility of the results \citep{Boylan2015-rm,Makridakis2018-sl}. For instance, this was true with the M4 competition where the vast majority of the methods were effectively reproduced by the organizers but not with the M5 where this was only done with the winners that were obliged to provide a clear description of their method along with their code, as well as a small number of the top 50 submissions that complied with the repeated requests of the organizers to share such information. 

Another idea would be for the organizers to make specific hypotheses before launching the competitions in order to test their predictions once the actual results become available, thus learning from their successes and mistakes. Such an approach would highlight the exact expectations of the competition and clarify its objectives, avoiding the problem of rationalizing the findings after the fact and allowing the equivalent of the scientific method, widely used in physical and life sciences, to be utilized in forecasting studies. This practice was followed in the M4 competition with positive results \citep{Makridakis2020-yw} and has been repeated with the M5. 

Finally, future forecasting competitions should challenge the findings of previous ones, testing the replicability of their results and trying to identify new, better forecasting practices as new, more accurate methods become available. For example, combining the forecasts of more than one methods has been a consistent finding of all competitions that has also flourished with machine and deep learning methods where ensembles of numerous individual models are used for producing the final forecasts. Another critical finding, lasting until the M4 competition, was that simple methods were at least as accurate as more sophisticated ones. This finding was reversed with the M4 and M5, as well as the latest Kaggle competitions, indicating the need for dynamic learning where new findings may reverse previous ones as new concepts (such as cross-learning) and more accurate methods are outperforming existing ones.

\section{Mapping the design attributes of past competitions}
\label{sec4}

In this section we map the design attributes discussed in section 3 to past, major forecasting competitions with the aim to identify their commonalities and design gaps, while also highlighting the advantages and drawbacks of each. We focus on the major competitions, organized by the community of the International Institute of Forecasters (IIF), but also on recent competitions hosted on Kaggle. In total, we consider seventeen forecasting competitions, which are listed in the rows of Tables \ref{tab:review1}, \ref{tab:review2}, and \ref{tab:review3} with the columns of the table presenting the various design attributes discussed in the previous section. Table \ref{tab:review4} offers citations to the relevant papers and links to the data and the winning submissions, where available. From the total of the seventeen competitions conducted in the last 40 years, seven were hosted by Kaggle, five were M competitions, three were energy ones, while there was a single tourism and a sole neural network one. 

\begin{landscape}

\begin{table*}[ht]
\scriptsize
\centering
  \caption{Mapping the design attributes of past competitions: Scope.}
    \begin{tabular}{p{0.15\linewidth}  p{0.15\linewidth} p{0.10\linewidth} p{0.10\linewidth} p{0.08\linewidth} p{0.08\linewidth} p{0.06\linewidth} p{0.08\linewidth} p{0.06\linewidth}}
    \hline
\textbf{Competition} & \multicolumn{8}{c}{\textbf{Scope}}\\
\cmidrule(lr){2-9}
\textbf{(Year)} & \multicolumn{3}{c}{\textbf{Focus}} & \multicolumn{2}{c}{\textbf{Type of submission}} & \multicolumn{3}{c}{\textbf{Format of submission}}\\
\cmidrule(lr){2-4}
\cmidrule(lr){5-6}
\cmidrule(lr){7-9}
&\textbf{Generic}&\textbf{Specific}&\textbf{Semi-specific}&\textbf{Numerical}&\textbf{Judgmental}&\textbf{Point forecasts}&\textbf{Uncertainty estimates}&\textbf{Decisions}\\
\hline
M or M1 (1982)&Macro, Micro, Industry \& Demographic&&&\cmark&&\cmark&&\\
\hline
M2 (1993)&Micro \& Macro&&&\cmark&\cmark&\cmark&&\\
\hline
M3 (2000)&Micro, Macro, Industry, Demographic, Finance \& Other&&&\cmark&&\cmark&&\\
\hline
NN3 (2006)&&&Industry&\cmark&&\cmark&&\\
\hline
Tourism (2011)&&&Tourism&\cmark&&\cmark&4 quantiles (computed only for benchmarks)&\\
\hline
GEFCom 2012 (2012)&&Load/Wind&&\cmark&&\cmark&&
\\
\hline
GEFCom 2014 (2014)&&&Load/ Wind/ Solar/ Price&\cmark&&&99 quantiles&\\
\hline
GEFCom 2017 (2017)&&Load&&\cmark&&&5 quantiles (9 in qualifying match)&\\
\hline
M4 (2018)&Micro, Finance, Macro, Industry, Demographic \& Other&&&\cmark&&\cmark&2 quantiles&\\
\hline
M5 (2020)&&Retail sales&&\cmark&&\cmark&9 quantiles&\\
\hline
Walmart Recruiting - Store Sales Forecasting (2014)&&Retail store sales&&\cmark&&\cmark&&\\
\hline
Walmart Recruiting II: Sales in Stormy Weather (2015)&&Retail sales of weather-sensitive products&&\cmark&&\cmark&&\\
\hline
Rossmann Store Sales (2015)&&Drug store sales&&\cmark&&\cmark&&\\
\hline
Grupo Bimbo Inventory Demand (2016)&&Bakery goods sales&&\cmark&&\cmark&&\\
\hline
Web Traffic Time Series Forecasting (2017)&&Traffic of web pages&&\cmark&&\cmark&&\\
\hline
Corporación Favorita Grocery Sales Forecasting (2018)&&Grocery store sales&&\cmark&&\cmark&&\\
\hline
Recruit Restaurant Visitor Forecasting (2018)&&Restaurant visits&&\cmark&&\cmark&&\\
\hline
    \end{tabular}
  \label{tab:review1}
\end{table*}

\begin{table*}[ht]
\scriptsize
\centering
  \caption{Mapping the design attributes of past competitions: Diversity and representativeness (specified based on the origin and size of the data set as well as the length of the period examined); Data structure; Data granularity; Data availability.}
    \begin{tabular}{p{0.15\linewidth}  p{0.12\linewidth} p{0.10\linewidth} p{0.10\linewidth} p{0.09\linewidth} p{0.09\linewidth} p{0.06\linewidth} p{0.15\linewidth}}
    \hline
\textbf{Competition} & \textbf{Diversity \&}  & \multicolumn{2}{c}{\textbf{Data structure}}  & \multicolumn{2}{c}{\textbf{Data granularity}}  & \multicolumn{2}{c}{\textbf{Data availability}} \\
\cmidrule(lr){3-4}
\cmidrule(lr){5-6}
\cmidrule(lr){7-8}
\textbf{(Year)} & \textbf{representativeness} & \textbf{Hierarchies} & \textbf{Exogenous variables} & \textbf{Cross
sectional} & \textbf{Temporal} & \textbf{Number of events} & \textbf{Observations per event (min-median-max)} \\
\hline
M or M1 (1982)&Moderate&&&From country to company&Monthly, Quarterly \& Yearly&1001&Monthly 30-66-132 Quart 10-40-106 Year 9-15-52\\
\hline
M2 (1993)&Low&&\cmark&Country and company&Monthly \& Quarterly&29&45-82-225 (Monthly) \& 167-167-167 (Quarterly)\\
\hline
M3 (2000)&Moderate&&&From country to company&Monthly, Quarterly, Yearly \& Other&3003&Monthly 48-115-126 Quart 16-44-64 Year 14-19-41 Others 63-63-96\\
\hline
NN3 (2006)&Low&&&From country to region&Monthly&111&50-116-126\\
\hline
Tourism (2011)&Moderate&&\cmark&From country to company&Yearly, Quarterly \& Monthly&1311&7-23-43 (Yearly), 22-102-122 (Quarterly) \& 67-306-309 (Monthly)\\
\hline
GEFCom 2012 (2012)&Moderate&Hierarchical / \xmark&\cmark (only for train data) / \cmark&Utility zone/Wind farm&Hourly&44397&38070/19033\\
\hline
GEFCom 2014 (2014)&Moderate&&\cmark&Utility/ Wind farm/Solar plant/Zone&Hourly&1/10/3/1&Round-based: 50376-60600/6576-16800/8760-18984/21528-25944\\
\hline
GEFCom 2017 (2017)&Moderate&Hierarchical&\cmark&Delivery point meters (zones in qualifying match)&Hourly&161 out of 169 (8 in qualifying match)&2232-61337 (Round based 119904-122736 in qualifying match)\\
\hline
M4 (2018)&High&&&From country to company&Monthly, Quarterly, Yearly, Daily, Hourly \& Weekly&100000&42-202-2794 (Monthly), 16-88-866 (Quarterly), 13-29-835 (Yearly), 93-2940-9919 (Daily), 700-960-960 (Hourly) \& 80-934-2597 (Weekly)\\
\hline
M5 (2020)&Moderate&Grouped&\cmark&Store-Product&Daily&30490&96-1782-1941\\
\hline
Walmart Recruiting - Store Sales Forecasting (2014)&Moderate&Grouped&\cmark&Store-Department&Weekly&3331&1-143-143\\
\hline
Walmart Recruiting II: Sales in Stormy Weather (2015)&Moderate&Grouped&\cmark&Store-Product&Daily&4995&851-914-1011\\
\hline
Rossmann Store Sales (2015)&Moderate&Grouped&\cmark&Store&Daily&1115&941-942-942\\
\hline
Grupo Bimbo Inventory Demand (2016)&Moderate&Hierarchical&&Store-Product&Weekly&26396648&1-2-7\\
\hline
Web Traffic Time Series Forecasting (2017)&Moderate&Grouped&&Page and Traffic Type&Daily&145063&803\\
\hline
Corporación Favorita Grocery Sales Forecasting (2018)&Moderate&Grouped&\cmark&Store-Product&Daily&174685&1-1687-1688\\
\hline
Recruit Restaurant Visitor Forecasting (2018)&Moderate&Grouped&\cmark&Restaurant&Daily&829&47-296-478\\
\hline
    \end{tabular}
  \label{tab:review2}
\end{table*}

\begin{table*}[ht]
\scriptsize
\centering
  \caption{Mapping the design attributes of past competitions: Forecasting horizon; Evaluation setup; Performance measurement; Benchmarking; Learning.}
    \begin{tabular}{p{0.15\linewidth}  p{0.13\linewidth} p{0.07\linewidth} p{0.07\linewidth} p{0.12\linewidth} p{0.05\linewidth} p{0.05\linewidth} p{0.15\linewidth} p{0.06\linewidth}}
    \hline
\textbf{Competition} & \textbf{Forecasting horizon}  & \multicolumn{2}{c}{\textbf{Evaluation setup}}  & \multicolumn{3}{c}{\textbf{Performance measurement}}  & \textbf{Benchmarks} & \textbf{Learning} \\
\cmidrule(lr){3-4}
\cmidrule(lr){5-7}
\textbf{(Year)} & \textbf{} & \textbf{Live} & \textbf{Rounds} & \textbf{Forecast} & \textbf{Utility} & \textbf{Cost} \\
\hline
M or M1 (1982)&1-18 (Monthly), 1-8 (Quarterly) \& 1-6 (Yearly)&&1&MAPE, MSE, AR, MdAPE \& PB&&&Naive \& ES&\cmark\\
\hline
M2 (1993)&1-15 (Monthly) \& 1-5 (Quarterly)&\cmark&2&MAPE&&&Naive, ES, ARIMA \& Combination of ES&\cmark\\
\hline
M3 (2000)&1-18 (Monthly), 1-8 (Quarterly), 1-6 (Yearly) \& 1-8 (Other)&&1&sMAPE&&&Naive, ES, Combination of ES \& ARIMA&\cmark\\
\hline
NN3 (2006)&1-18&&1&sMAPE&&&Naive, ES, Theta, Combination of ES, Expert systems, ARIMA, vanilla NNs and SVR&\cmark\\
\hline
Tourism (2011)&1-4 (Yearly), 1-8 (Quarterly) \& 1-24 (Monthly)&&1&MASE \& Coverage&&&Naive, ES, Theta, ARIMA, Expert systems \& models with explanatory variables&\cmark\\
\hline
GEFCom 2012 (2012)&1-168/1-48&&1/1 of 157 periods&RMSE&&&Vanilla MLR/Naive&\cmark\\
\hline
GEFCom 2014 (2014)&1-24 for Price \& (1-31)*24 for the rest&&15&PL improvement over benchmark, adjusted for simplicity and quality&&&Naive&\cmark\\
\hline
GEFCom 2017 (2017)&8784 ((1-31)*24 in qualifying match)&\cmark&1 (6 in qualifying match)&PL improvement over benchmark&&&Vanilla MLR&\cmark\\
\hline
M4 (2018)&1-18 (Monthly), 1-8 (Quarterly), 1-6 (Yearly), 1-14 (Daily), 1-48 (Hourly) \& 1-13 (Weekly)&&1&OWA \& MSIS&&&Naive, ES, Theta, Combination of ES, ARIMA \& vanilla NNs&\cmark\\
\hline
M5 (2020)&1-28&&1&WRMSSE/WSPL&&&Naive, ES, ARIMA, Croston and variants, Combinations, NNs, RTs&\cmark\\
\hline
Walmart Recruiting - Store Sales Forecasting (2014)&1-39&&1&WMAE&&&All zeros&\\
\hline
Walmart Recruiting II: Sales in Stormy Weather (2015)&1-25 (in an interpolation fashion)&&1&RMSLE&&&All zeros&\\
\hline
Rossmann Store Sales (2015)&1-48&&1&RMSPE&&&All zeros, Median day of week&\\
\hline
Grupo Bimbo Inventory Demand (2016)&1-2&&1&RMSLE&&&All sevens&\\
\hline
Web Traffic Time Series Forecasting (2017)&3-65&\cmark&1&sMAPE&&&All zeros&\\
\hline
Corporación Favorita Grocery Sales Forecasting (2018)&1-16&&1&NWRMSLE&&&Mean item sales, Last year sales, All zeros&\\
\hline
Recruit Restaurant Visitor Forecasting (2018)&1-39&&1&RMSLE&&&Median visit, All zeros&\\
\hline
    \end{tabular}
  \label{tab:review3}
\end{table*}

\end{landscape}

\begin{table*}[ht]
\scriptsize
\centering
  \caption{Past forecasting competitions: Citations and additional information.}
\begin{tabular}{p{0.20\linewidth}  p{0.25\linewidth} p{0.50\linewidth}}
\hline
\textbf{Competition} & \textbf{Citation}  & \textbf{Links to data and/or winning methods} \\
\hline
M or M1 (1982)&\cite{Makridakis1982-co} & https://forecasters.org/resources/time-series-data/ https://cran.r-project.org/package=Mcomp\\
\hline
M2 (1993)&\cite{Makridakis1993-bw} & https://forecasters.org/resources/time-series-data/\\
\hline
M3 (2000)&\cite{Makridakis2000-ty} & https://forecasters.org/resources/time-series-data/ https://cran.r-project.org/package=Mcomp\\
\hline
NN3 (2006)&\cite{Crone2011-si} & http://www.neural-forecasting-competition.com/NN3\\
\hline
Tourism (2011)&\cite{Athanasopoulos2011-lw}&https://www.kaggle.com/c/tourism1 https://www.kaggle.com/c/tourism2 https://github.com/robjhyndman/tscompdata\\
\hline
GEFCom 2012 (2012)&\cite{Hong2014-sz} & https://www.kaggle.com/c/global-energy-forecasting-competition-2012-load-forecasting http://www.drhongtao.com/gefcom/2012\\
\hline
GEFCom 2014 (2014)&\cite{Hong2016-my}&http://www.drhongtao.com/gefcom/2014\\
\hline
GEFCom 2017 (2017)&\cite{Hong2019-ay}&http://www.drhongtao.com/gefcom/2017\\
\hline
M4 (2018)&\cite{Makridakis2020-mm}&https://forecasters.org/resources/time-series-data/ https://github.com/Mcompetitions/M4-methods https://github.com/carlanetto/M4comp2018\\
\hline
M5 (2020)&\cite{Makridakis2020-wq,Makridakis2020-bj}&https://www.kaggle.com/c/m5-forecasting-accuracy https://www.kaggle.com/c/m5-forecasting-uncertainty https://forecasters.org/resources/time-series-data/ https://github.com/Mcompetitions/M5-methods\\
\hline
Walmart Recruiting - Store Sales Forecasting (2014)&&https://www.kaggle.com/c/walmart-recruiting-store-sales-forecasting\\
\hline
Walmart Recruiting II: Sales in Stormy Weather (2015)&&https://www.kaggle.com/c/walmart-recruiting-sales-in-stormy-weather\\
\hline
Rossmann Store Sales (2015)&&https://www.kaggle.com/c/rossmann-store-sales\\
\hline
Grupo Bimbo Inventory Demand (2016)&&https://www.kaggle.com/c/grupo-bimbo-inventory-demand\\
\hline
Web Traffic Time Series Forecasting (2017)&&https://www.kaggle.com/c/web-traffic-time-series-forecasting\\
\hline
Corporación Favorita Grocery Sales Forecasting (2018)&&https://www.kaggle.com/c/favorita-grocery-sales-forecasting\\
\hline
Recruit Restaurant Visitor Forecasting (2018)&&https://www.kaggle.com/c/recruit-restaurant-visitor-forecasting\\
\hline
    \end{tabular}
  \label{tab:review4}
\end{table*}

There are several common attributes characterizing practically all seventeen competitions. First, the submissions required were all numerical except for M2 that asked, in addition, for judgmental inputs from the forecasters. Second, in all but three competitions (M2, GEF2012, and GEF2017) the submission setup involved a fixed origin evaluation on concealed data. Third, there were only three live competitions (M2, GEF2017, and Web Traffic Time Series Forecasting) that were also limited in a small number of evaluation rounds. Fourth, the majority of the competitions (fifteen out of the seventeen) required point forecasts while five also demanded uncertainty estimates, ranging from 2 quantiles in M4 to 99 in GEF2014. Fifth, while there is a balance between generic, specific, and semi-specific competitions, we observe that specific ones focus on tourism, energy, and retail forecasting applications, with the majority of the specific ones including high-frequency, hierarchically structured series and explanatory/exogenous variables, while the generic ones focusing on lower-frequency data, such as yearly, quarterly, and monthly, that were not accompanied with additional information. Moreover, there seems to be a trend towards more detailed data sets as more recent competitions move from individual time series to hierarchically structured ones that may be influenced by explanatory/exogenous variables. Sixth, none of the competitions required submissions in the form of decisions nor evaluated their performance in terms of utility or cost-based measures, utilizing various statistical measures that build on absolute, squared, and percentage errors. Finally, with the exception of the competitions that were organized by academics, little emphasis was given on the element of learning and how to improve forecasting performance and few, non-competitive benchmarks were considered for evaluating such improvements. For instance, the M and energy competitions included several variations of naive approaches, combinations of exponential smoothing models, ARIMA, the Theta method, and simple machine learning or statistical regression methods, while Kaggle ones featured only naive methods and dummy submissions (e.g. all forecasts are set equal to the global average/median or zero).

These observations reveal both a consensus to apply what has worked in the past and it is easy to implement in practice as well as a desire for experimentation. What is clear from Tables \ref{tab:review1}, \ref{tab:review2}, and \ref{tab:review3} is the difference between the top twelve competitions organized by the academic community and the last five ones hosted by Kaggle. In the former the emphasis is on learning by publishing the results in peer reviewed journals and providing open access to the data and forecasts so that others can comment on the findings, respond to their value, and suggest improvements for future ones. Thus, it is not surprising that the number of citations received by the former (close to 5,500) are significantly more than that of the latter (probably limited to less than 100). Citations are an integral part of learning as other researchers read the cited work and become aware of its findings that they then try to extend to additional directions. At the same time, the Kaggle approach encourages cooperation among competitors, e.g. in the form of forum discussions and code exchange, to come up with the best solution to the problem at hand without concern with the dissemination of the findings to the wider data science community. Equally important, Kaggle involves public leader-boards that provide instant feedback to the participants in order for them to revise their methods and resubmit forecasts, thereby encouraging competition and driving innovation \citep{ATHANASOPOULOS2011845}. A clear breakthrough will come by combining the academic and Kaggle approaches by exploiting the advantages of both as there is no reason that Kaggle scientists will not be willing to share their knowledge so that others can also learn from their experience, nor for the academics not to be benefited by leader-boards, public discussions, or code exchange. In our view, such a breakthrough will be inevitable to happen in the near future.

\section{Future forecasting competitions}
\label{sec5}

\subsection{Proposed principles}\label{sec5-1}
In the previous sections we discussed the design aspects of forecasting competitions and mapped these to the past ones. Then, we elaborated on the design opportunities, i.e., the gaps that past forecasting competitions have left. In this section, we propose some principles for future competitions:

\textbf{Replicability.} 
One crucial aspect of any research study is that its results should be able to be replicated. This has been an increasing concern across sciences \citep{Goodman2016-wn}, including the forecasting field \citep{Boylan2015-rm,Makridakis2018-sl}. To be most useful to the forecasting community, competitions should ideally be transparent and allow for the full replicability of the results. One way to achieve this is by requiring submission of the source code (or at least an executable file) of the participating solutions, coupled with sufficient descriptions and open libraries for benchmarks and performance measures. Reproducibility will also allow those interested to test if the results of a forecasting competition hold for other data sets, performance measures, forecasting horizons, and testing periods, while also enabling computational cost comparisons. In addition it would enable rolling-origin evaluation to be done in an automated fashion, reflecting the realistic situation when forecasting models are built and then run repeatedly without the opportunity to tweak them each time an output is generated.

\textbf{Representativeness.} 
If possible, organizers of forecasting competitions should aim for a diverse and representative set of data. A high degree of representativeness \citep{Spiliotis2020-gl} will allow for a fuller analysis of the results, enabling us to understand the conditions under which some methods perform better than others. Moving away from ``the overall top-performing solution wins it all'', we will be able to effectively understand the importance of particular features (including frequencies) and gain insights of the performance of various methods for specific industries or organizations. One strategy to improve representativeness could be to look at the feature space for time series included in a competition in comparison with other samples from the relevant population of series \citep{Kang2017-ha,Spiliotis2020-gl,Fry2020-ye}. 

Forecasting under highly stable conditions offers little challenge. Therefore, competition organizers should consider evaluating forecasts across a range of conditions, including conditions where past patterns/relationships are bound to fail (e.g., structural changes, fat tails, recessions, pandemics) to identify methods that are more robusts under such conditions in order to offer valuable insights and enhance our understanding towards managing such situations. Moreover, including competitive actions and reactions should be included as this is the reality modern companies operate. Future competitions could also explore the possibility of multivariate (but not hierarchically structured) sets of data that also include information directly coming from online data devices, including nowcasting.

\textbf{Robust evaluation.} 
For the results of a competition to be meaningful, robust evaluation strategies must be considered. We suggest moving away from evaluating forecasts produced from a single origin, especially when the data set considered is homogeneous, and introduce rolling evaluation schemes. This would be particularly relevant for seasonal time series, where evaluation periods should cover many different times within the calendar year, if not one or more complete years.
This would mitigate against sampling bias caused by evaluating forecasts over a short interval. Organizers could also consider evaluating hold-out sets for representativeness using the principles discussed above. Future competitions could also offer multiple evaluation rounds in a live setup. Undoubtedly, this would add a more pragmatic dimension to forecasting competitions.

\textbf{Measuring impact on decisions.}
Reflection to reality may include how a forecasting solution is indeed implemented in practice, but also offer metrics of performance measurement that are directly linked to decisions. For example, in inventory forecasting, \cite{Petropoulos2019-bu} map the forecasting performance of various forecasting methods to their inventory performance, measured in terms of holding cost, achieved service levels, and variance of forecasts. We argue that future forecasting competitions may need to shift the focus to measure the utility of the forecasts/uncertainty directly. In many applications, the translation from point and probabilistic forecasts to their decision making implications is a big step and a formidable challenge as utility can be not only non-linear but also non-monotonic. Whenever possible, such utility should be expressed in monetary terms that would allow comparing meaningful trade-offs. Such trade-offs could include conflicting optimization criteria (such as inventory holdings versus service levels) but also would allow for a more systematic value-added analysis of the complexity of the participating solutions and their computational (or otherwise) cost. However, we should be careful to distinguish between evaluating the impact of forecasts on decisions versus evaluating the impact of decisions themselves.

\textbf{Showcase forecast-value-added (FVA). }
Forecasting competitions need to clearly demonstrate the added-value of a proposed solution over the state-of-the-art methods and benchmarks. The choices for benchmarks is wide and could include top-performing methods from previous competitions. For instance, a future large-scale generic forecasting competition could have as a benchmark the winning method of \cite{Smyl2020-dv} in the M4 competitions, or N-Beats \citep{oreshkin2020nbeats}. Also, a future competition on retail forecasting should include as benchmarks the top-methods from M5 or other Kaggle competitions. Finally, the inclusion of past winning approaches as benchmarks can act as a way of measuring improvements from new competitions and determining the value they have added in forecasting performance. We suggest that an FVA analysis should be multifold and include not only the performance of the point forecasts, but also the performance in estimating uncertainty, dealing with fat tails, and the computational cost and complexity of each method. The last two aspects (complexity and cost) are increasingly important for the acceptance and successful implementation of a method particularly when millions of forecasts/estimates of uncertainty are needed on a weekly basis.

\textbf{Enhancing knowledge.} 
We would like to see future competitions focus on contributing new learnings and insights to the forecasting community, moving away from a horse-race exercise towards bridging the gap between theory and practice. If possible, forecasting competitions should not focus on picking a winning team, but rather understanding what constitutes a winning method, i.e., a successful underlying mechanism, and how the results could be transferable to other settings. For instance, cross-learning has been proven to be an effective method in the M5 competition set-up where grouped series were forecast, while combinations is a winning approach for univariate time series forecasting tasks. They should be able to show how the results can be implemented to improve the baseline and what are the consequences of the forecasting accuracy/uncertainty on decision making. Although not an objective of all past forecasting competitions, learning must become an integral part of all future ones to maximize their expected value by making their findings widely known to anyone wishing to utilize them to improve the theory or practice of forecasting. The current trend towards open access of knowledge must be applied to forecasting competitions as its findings will improve the much talked circular economy by eliminating waste and achieving optimal results across a wide variety of operational and strategic areas.

\textbf{Merging the academic and Kaggle approaches.} 
There is much to gain and nothing to lose by combining the academic approach of disseminating learning and achieving high citations with that of Kaggle encouraging high collaboration and open participation by the participating groups. Facilitating learning by widely disseminating the findings of Kaggle competitions will benefit the entire data science community and avoid concerns about their relevance \citep[see][]{Chawla2020-ce}. At the same time, stimulating a more supportive collaborative spirit in academic competitions can encourage innovation and foster team effort, as long as some clever ways of supporting collaborative work could be adopted. 

We note that one strategy that enables both replicability and robust evaluation is the use of code-only competitions, where the organizers of the competition use the submitted codes to produce forecasts for multiple origins. Participants may be given the option to alter their code in key points; for instance, the participants may resubmit their codes every quarter when forecasts are produced and evaluated every week. Such a strategy also reflects the real-world situation in that a forecasting model used in practice may not be able to benefit from manual tweaking between each subsequent forecast generation, leading also to unreasonably higher costs in terms of post performance analysis and re-engineering. In a code-only competition, an additional requirement could be that the code must run within a specific time limit, given a specific data set and a computer architecture. That would focus attention on finding methods that achieve optimal outcomes within a computationally constrained environment, limiting the need to consider metrics related to the computational cost.

\subsection{Towards forecasting athlons}\label{sec5-2}
The capabilities of forecasts to make and use forecasts have progressed significantly during the last four decades, based in part on the findings of forecasting competitions that as \cite{Hyndman2020-gs} mentions have contributed a great deal to improve the theory and practice of forecasting and provide considerable value to business firms using such predictions to improve their operations. Forecasting competitions could be further expanded beyond business applications to other social science areas to provide objective information and improve policy and decision making. In addition, uncertainty needs to receive attention among academicians and practitioners alike. It must be accepted that uncertainty will always exist and cannot be avoided or reduced no matter if we would like to live in a world without uncertainty. What we will have to do is to understand its risk implications and consider what actions to take to minimize the negative consequences involved. Directly linking forecasting competitions with decision-making aspects and the utility of the forecasts is also very important and will allow us to gain further insights on the use of forecasts in practice. 

Attempting to incorporate all of these consideration into a single forecasting contest or evaluation can be difficult, if not impossible. Therefore, we suggest that some future forecasting competitions could move from featuring a single challenge to multiple ones, with a winner in each challenge and an overall winner for the entire competition. For example, a future forecasting competition could be a pentathlon (or hexathlon, or heptathlon...), where the various challenges could be organized around domain skills, such as (\textit{i}) forecasting of univariate series with no exogenous information, (\textit{ii}) forecasting of multivariate series, (\textit{iii}) forecasting of series with exogenous information (e.g., weather, price, promotion activity, competitor actions, etc.), (\textit{iv}) long-range forecasting with market or competitor uncertainties, (\textit{v}) forecasting of intermittent series, (\textit{vi}) lifecycle forecasting, etc. We view this structure as valuable for a comprehensive forecasting competition for several reasons:

First, as noted above, it may be impossible to cover all of the ideal aspects and core forecasting skills in a single challenge. Second, the use of multiple challenges would also allow for greater diversity of application domains. And third, this would enable evaluation of participants in multiple skill domains and would reduce the randomness in the final results and rankings.

Another possibility would be the organization of challenges around applications. For example, within a manufacturing company, that could include forecasting for (\textit{i}) inventory, (\textit{ii}) scheduling, (\textit{iii}) budget, (\textit{iv}) cash flows, (\textit{v}) long-range planning, and (\textit{vi}) human resources, among others. Such challenges will better reflect reality and showcase FVA since, in real life, in order for an organization to thrive, accurate forecasts and correct estimates of uncertainty are required for multiple aspects of its strategy, planning, and operation related decisions.

Future domain-specific competitions could focus on new application areas, covering the economy (gross domestic product, monetary policies, interest rates), finance (stocks, commodities), operations (new products, promotional forecasting, spare parts, predictive maintenance, reverse logistics), healthcare (epidemics, healthcare management, mortality, preventable medical errors), climate, sports, elections, call centers, big projects and megaprojects, transportation, and online commerce, among others. Finally, the increasing role of judgment in various aspects of the forecasting process, such as adjusting/finalizing forecasts or even selecting between models, calls for further investigations and its should be further explored in future competitions. 

Overall, we foresee that forecasting competitions have still much to offer if they are designed in a way to represent reality even closer. If forecasting competitions are done systematically and consistently, they would allow for comparisons and assessing improvements over time while covering also various areas of applications and time horizons.

\section{Conclusions}
\label{sec6}

Forecasting competitions, the equivalent of laboratory experimentation in physical and life sciences, provide useful, objective information to improve the theory and practice of forecasting, advancing the field and enhancing decision and policy making. This paper has described all major past forecasting competitions, discussed their design attributes, and identified those of ``ideal'' competitions, extending their coverage to a multitude of applications and social science areas, echoing Hyndman’s suggestion that the main objective of competitions is learning as much as possible rather than identifying winners. 

The main part of the paper described ten design attributes to be considered by the organizers of competitions who need to decide those relevant for their own, considering trade-offs between optimal choices and practical concerns like costs, as 
well as elements related with the time and effort required to participate in them. Next, the paper mapped all pertinent past competitions in respect of the described design attributes, identifying similarities and differences between the competitions, as well as design gaps, and making suggestions about the attributes that future competitions should consider, putting a particular emphasis on learning as much as possible from their implementation in order to help improve forecasting accuracy and uncertainty.

The majority of past competitions concentrated on point forecasts. Our proposal is that all future competitions should also request probabilistic forecasts for a sufficient number of quantiles so that both the main part of the uncertainty distribution and its tails are effectively captured. This is of critical importance since both point forecasts and uncertainty estimates need to be considered in all future oriented decisions. Another concentration of past competitions is the usage of the single origin concealed data evaluation setup, which is the easiest to implement and requires the least time to participate. This practice will have to change by first expanding the evaluation setup to several rolling origins and then potentially moving to rolling live competitions that may be the hardest to run but provide a great value as they run on a real-time basis where all information is currently available and judgmental inputs can be directly incorporated. Clearly, there will be trade-offs that would need to be considered between the number of rolling origins used and the amount of effort that would be required to complete the competition, with the same trade-offs deliberated between live and concealed data ones. Competitions are costly to run, requiring a considerable amount of effort both to be implemented and participate. Their advantage is the objective evidence they provide to improve the theory and practice of forecasting. As such, they must continue and maybe their costs are financed by a joint industry or specific group effort in search of solutions to improve the accuracy and uncertainty of their specific predictions. Whatever the solution, the practice of forecasting competitions must expand in the future to gain the maximum benefits from their findings.

The final section of the paper ends with the observation that the task of forecasting presents a multitude of challenges for organizations and societies. Business firms, for instance, must predict the level of their inventories for the large number of items sold in their stores, schedule their production and workforce, and estimate their budget requirements as well as their long term strategic plans, including competitive and technological forecasts. Moreover, economic forecasting is also necessary at the societal level as well as energy, climate, and health predictions. Such multitude of challenges cannot be met with a single competition. Instead, a number of them would be demanded like in a pentathlon where different challenges take place, identifying the winner of each but also the overall one that would contribute the most to the overall forecasting effort among various areas or even industries with varying characteristics.


\singlespacing

\bibliographystyle{elsarticle-harv}
\bibliography{refs}

\end{document}